\documentclass[%
reprint,
nofootinbib,
 amsmath,amssymb,
 aps,
]{revtex4-2}

\usepackage{physics}
\usepackage{braket}
\usepackage{graphicx}
\usepackage{dcolumn}
\usepackage{bm}
\usepackage{float}
\usepackage{xcolor}

\begin{document}

\preprint{APS/123-QED}

\title{{Integrated photonic sources of frequency-bin-encoded multipartite entangled states}}

\author{Milica Banic$^1$}
\email{mbanic@physics.utoronto.ca}
\author{J. E. Sipe$^1$}
\author{Marco Liscidini$^2$}
\affiliation{$^1$ Department of Physics, University of Toronto, 60 St. George Street, Toronto, ON, M5S 1A7, Canada\\ $^2$ Department of Physics, University of Pavia, Via Bassi 6, 27100, Pavia, Italy}

\date{\today}

\begin{abstract}
{We demonstrate that genuine multipartite entangled states can be generated using frequency bin encoding in integrated photonic platforms.} We introduce a source of four-photon GHZ states, and a source of three-photon W states. We predict generation rates on the order of 10$^4$ Hz for a silicon microring source with milliwatt pump powers. {These results, along with the versatility and scalability of integrated structures, identify this as a promising approach for the generation of higher-dimensional and larger entangled states.}
\end{abstract}

\maketitle


\section{Introduction}
\label{section:introduction}
Entangled states are an indispensible resource for a host of quantum protocols, and for tests of fundamental physics. There is now an abundance of photonic sources of bipartite entangled states, and strategies for generating multipartite states are being explored \cite{QD_GHZme}. Multipartite entangled states, which exist in a Hilbert space with three or more factor spaces, are particularly interesting from both a fundamental and practical point of view: They exhibit correlations that cannot be reproduced with only bipartite states. 

Multipartite states can be encoded in a number of ways, including using multiple degrees of freedom in only two particles \cite{Imany2019}. However, many applications rely on \emph{genuine} multipartite entangled states \cite{PhysRevA.59.1829}; we define genuine multipartite states for which the Hilbert state is composed of $n$ factor spaces which correspond to $n$ physically separable entities, for example, $n$ photons.

Genuine multipartite states can be composed of photons generated by parametric nonlinear processes, such as spontaneous four-wave mixing (SFWM) or spontaneous parametric down-conversion. The logical states can be encoded in a number of ways. Many approaches make use of polarization encoding in bulk systems \cite{PhysRevLett.86.4435, Silberhorn_PhysRevLett.129.150501}, but this scheme has drawbacks: It is constrained to two logical states per particle -- e.g., horizontal or vertical polarization -- and it lacks scalability, since polarization is difficult to control in integrated systems.

Motivated by the need for integration, path encoding has also been explored. In this scheme, the logical state is encoded by the waveguide in which the photon is detected \cite{Bergamasco_PRApp, Li:22}. Although this can be implemented on-chip \cite{Bao2023}, its scalability is challenging because increasing the dimensionality of the system requires increasing the number of waveguides. In some cases, the implementation of large states is also complicated by the need to avoid waveguide crossings, at least for conventional lithography processes. 

Another approach is energy encoding, where information is encoded in the photons' frequencies. This degree of freedom is scalable, robust, and compatible with integrated platforms. Taking photons generated by SFWM, one possibility is to encode the logical states in the photons' colour: Signal (red with respect to the pump) or idler (blue). This scheme has been implemented in bulk for the generation of energy-encoded W states \cite{Fang_PRL}, and it can in principle be extended to integrated structures \cite{Menotti_PRA}. However, the relatively large spacing between the logical states' frequencies makes their manipulation challenging.  

\begin{figure}[h]
    \centering
    \includegraphics[width=0.3\textwidth]{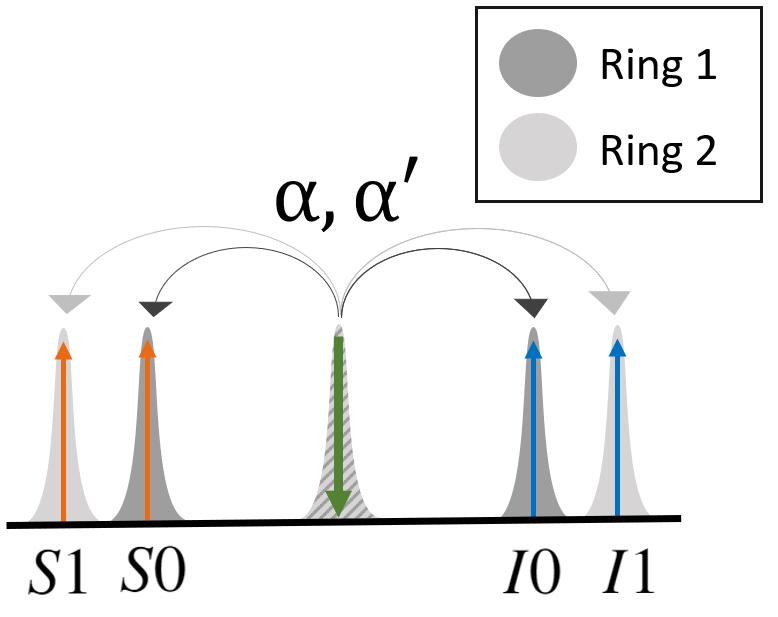}
    \caption{Pumping scheme described by Eq. \eqref{eq:HNL_00_11}. Green arrows represent classical pumps (with amplitudes $\alpha$, $\alpha'$); orange and blue arrows represent signal and idler fields. {$S0$, $S1$, $I0$, $I1$} label the ring resonances.}
    \label{fig:single_pump}
\end{figure}

Recently, frequency bin encoding has been explored. This is a type of energy encoding in which the frequencies are close enough to be manipulated using commercial electro-optic modulators \cite{Imany2018}. It has been demonstrated that more than two logical states could be implemented in this approach, making frequency bin encoding a promising candidate for the generation of qudit
states \cite{Imany2018, doi:10.1126/science.aad8532_Reimer}. The generation of frequency-bin-encoded states can be implemented using photons generated by SFWM in a resonator, in which the photon pairs are generated in a comb of resonances \cite{Kues2017,Liscidini:19,Clementi2023}. In this scheme, the logical state of a photon is determined by {its spectral distance from the pump resonance} 
(see Fig. \ref{fig:single_pump}).

The generation of frequency-bin-encoded bipartite entangled states, such as Bell states, has already been demonstrated \cite{PhysRevLett.129.230505}; in the present work, we discuss the generation of frequency-bin-encoded multipartite states in integrated photonic devices. We demonstrate how this strategy can be used for the generation of W states and GHZ states, two paradigmatic examples of multipartite entangled states. In Section \ref{section:sources} we describe the multi-photon sources employed in these devices. In Section \ref{section:GHZ} we discuss the generation of four-photon GHZ states, and in Section \ref{section:W} we do the same for three-photon W states. Finally, we draw our conclusions in Section \ref{section:Conclusions}.

\section{Photon sources}
\label{section:sources}
The schemes to be discussed begin with the generation of entangled photons by spontaneous four-wave mixing (SFWM) in ring resonators. {We assume that the sources generate uncorrelated photon pairs; this can be accomplished by driving the microring source with appropriately shaped pulses \cite{Christensen:18}, or by using more complex ring resonator structures as the source \cite{Vernon:17}. Here we will assume the sources are driven by a train of pulses with a duration shorter than the dwelling time of the ring resonators. Under these conditions the generated photon pairs are nearly uncorrelated \cite{Grassani2016, Helt:10}, and to good approximation the photons are generated in a single Schmidt mode.}

We consider two configurations for the pump and generated modes: In the first, a pump field centered at a single resonance frequency is used to generate photons in two pairs of ring resonances.
%
{Taking the generated pairs to be uncorrelated, the nonlinear Hamiltonian for this pumping  scheme can be written as}
\begin{align}
    H_{NL}(t) = \hbar \Gamma \alpha^2(t) a^{\dagger}_{S0} a^{\dagger}_{I0} + \hbar \Gamma' \alpha'^2(t) a^{\dagger}_{S1} a^{\dagger}_{I1} + H.c., \label{eq:HNL_00_11}
\end{align}
where $\Gamma$ and $\Gamma'$ are nonlinear coupling rates, $\alpha(t)$ and $\alpha'(t)$ are pump amplitudes, and $a^{\dagger}_J$ is the raising operator associated with the resonance labelled by $J$. Each resonance has two labels associated with it: a \emph{frequency bin} (0 or 1) and a `\emph{colour}' with respect to the pump {($S$ or $I$)}. Although the logical state of a photon is encoded in the frequency bin, which depends only on its {spectral distance} from the pump, the photons also have a redundant label that specifies whether they are red- or blue-shifted with respect to the pump. Here energy conservation ensures that pairs of signal and idler photons are generated in the same frequency bin.

Eq. \eqref{eq:HNL_00_11} can be implemented using a single ring, provided the free spectral range is sufficiently small that the photons' frequency bins can be modulated using commercial electro-optic modulators. Another approach is the use of two rings with different radii \cite{Liscidini:19}, such that each is driven by the same pump field, with one ring generating photon pairs in the resonances $S0$ and $I0$, while the other generates photon pairs in $S1$ and $I1$ frequencies. If two rings are used, the distance between the frequency bins does not depend on the rings' FSR, so the frequency bins can spectrally close without affecting the generation rate, which scales quadratically with the FSR \cite{Helt:JOSAB}. The two rings could also be driven by a different pump amplitude and phase, hence the distinction between $\alpha(t)$ and $\alpha'(t)$ in Eq. \eqref{eq:HNL_00_11}.

\begin{figure}
    \centering
    \includegraphics[width=0.3\textwidth]{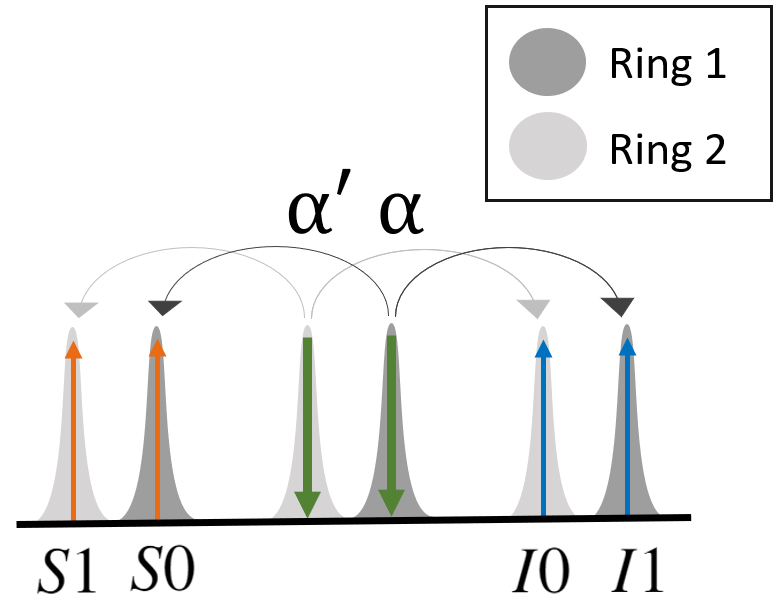}
    \caption{Dual pumping scheme described by Eq. \eqref{eq:HNL_01_10}. Green arrows represent classical pumps (with amplitudes $\alpha$, $\alpha'$); orange and blue arrows represent signal and idler fields. {$S0$, $S1$, $I0$, $I1$} label the ring resonances.}
    \label{fig:Dual_pumping}
\end{figure}

We will also consider a dual-pump scheme (see Fig. \ref{fig:Dual_pumping}), such that the generation of photons is described by 
\begin{align}
    H_{NL} = \hbar \Gamma \alpha^2(t) a^{\dagger}_{S0} a^{\dagger}_{I1} + \hbar \Gamma' \alpha'^2(t) a^{\dagger}_{S1} a^{\dagger}_{I0} + H.c.,\label{eq:HNL_01_10}
\end{align}
{where again we have assumed the photon pairs are approximately separable.} Here energy conservation requires that one photon of each pair is generated in frequency bin 1 while the other is generated in bin 0. Unlike in the previous configuration, $\alpha$ and $\alpha'$ refer to pump amplitudes at different frequencies, so they could be made distinct even in an implementation with a single ring. However, two high-finesse rings could be used to increase the generation rate \cite{Clementi2023}.

Neglecting time ordering corrections, the state generated in either pumping scheme can be written as 
\begin{align}
    \ket{\psi} &= e^{-\frac{i}{\hbar}\int^t_{t_0} dt' H_{NL}(t')} \ket{\text{vac}}. \label{eq:SE}
\end{align}
Inserting \eqref{eq:HNL_00_11} or \eqref{eq:HNL_01_10}, we can write 
\begin{align}
    \ket{\psi} &= e^{\beta C^{\dagger}_{II} - H.c.}\ket{\text{vac}},
    \label{eq:squeezed_state}
\end{align}
where $C^{\dagger}_{II}$ is a pair generation operator. We will consider the parameter regime where to good approximation, at most four photons are generated simultaneously, with the probability of higher-order events being negligible. In this case \eqref{eq:squeezed_state} can be approximated as
\begin{align}
    \ket{\psi} &\approx \left(1+\mathcal{O}(\beta^2)\right)\ket{\text{vac}} + \beta C^{\dagger}_{II}\ket{\text{vac}}+\frac{\beta^2}{2}C^{\dagger2}_{II} \ket{\text{vac}},
    \label{eq:ket_approx}
\end{align}
where $C^{\dagger}_{II}\ket{\text{vac}}$ is a normalized two-photon state, and $|\beta|^2$ is the probability of generating a photon pair per pump pulse, in the low pair generation regime \cite{Onodera_PhysRevA.93.043837, Bergamasco_PRApp, Menotti_PRA}.

\section{Four-photon GHZ states}
\label{section:GHZ}

We first discuss the generation of frequency-bin-encoded four-photon GHZ states. A maximally entangled four-qubit GHZ state has the form 
\begin{align}
    \ket{GHZ} = \frac{1}{\sqrt{2}}\left( \ket{0000} + \ket{1111}\right),
\end{align}
where 0 and 1 denote the qubits' logical states. GHZ states are resources for tasks ranging from fundamental tests of quantum mechanics to applications in quantum information and communication
\cite{Pan2000_GHZ, PhysRevA.63.054301_GHZ_dense_coding,PhysRevA.59.1829_GHZ_secret_sharing}.

\begin{figure}[h]
    \centering
    \includegraphics[width=0.5\textwidth]{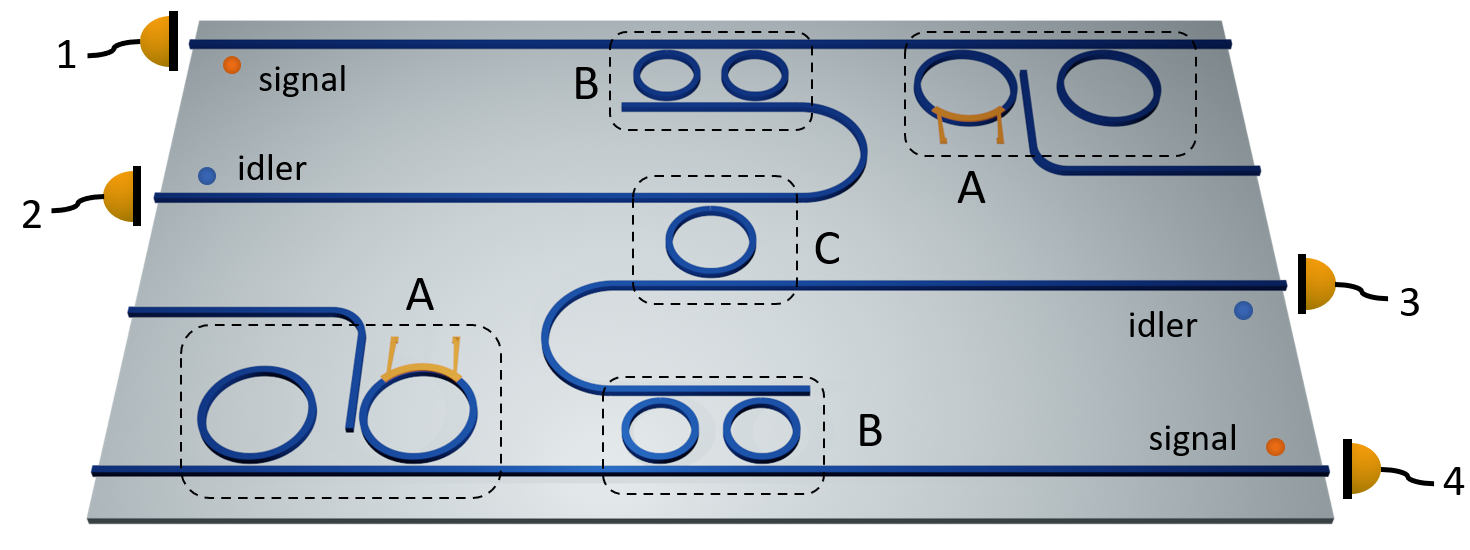}
    \caption{Sketch of the GHZ state source. Box A labels the photon pair sources. The generated signal and idler photons are separated by demultiplexers (labelled B). C labels an add-drop filter resonant with frequency bin I1.}
    \label{fig:GHZ_device}
\end{figure}

The source of frequency-bin-entangled GHZ states is sketched in Fig. \ref{fig:GHZ_device}. The scheme implemented by this device is analogous to the one implemented in bulk systems for the generation of polarization-encoded GHZ states \cite{PhysRevLett.86.4435}, and it relies on postselection on four-fold coincidences. 

In our implementation, we begin with the generation of photon pairs in two microring sources in the single-pump configuration (labelled `A' in Fig. \ref{fig:single_pump}). {We assume all four rings are driven simultaneously to generate photons that are indistinguishable in time; in Fig. \ref{fig:GHZ_device} we envision splitting the pump such that each ring is coupled to its own input waveguide}. 

The state generated by each source is described by Eq. \eqref{eq:ket_approx}. Inserting Eq. \eqref{eq:HNL_00_11} in \eqref{eq:SE} we have 
\begin{align}
    C^{\dagger}_{II} = \frac{1}{\beta}\left\{ \beta_1 a^{\dagger}_{S0} a^{\dagger}_{I0} + \beta_2 a^{\dagger}_{S1} a^{\dagger}_{I1}\right\},
\end{align}
where
\begin{align}
    \beta_1 &= -i\Gamma \int_{t_0}^{t} \alpha^2 (t') dt' \label{eq:beta1}\\
    \beta_2 &= -i\Gamma' \int_{t_0}^{t} \alpha'^2 (t') dt' \label{eq:beta2}
\end{align}
are related to the probabilities of generating a pair of photons in each pair of resonances (S0 and I0, or S1 and I1), and $|\beta|^2 = |\beta_1|^2 + |\beta_2|^2$ is again the total probability of generating a photon pair per pump pulse. The state generated by each source then is
\begin{align}
    \ket{\psi} = \ket{\text{vac}} + \left(\beta_1 a^{\dagger}_{S0} a^{\dagger}_{I0} + \beta_2 a^{\dagger}_{S1} a^{\dagger}_{I1}\right)\ket{\text{vac}},
\end{align}
{where have restricted our attention to the first-order term. In principle, four photons can be generated in a single source. However, our postselection on signal photons prevents such events from causing four-fold coincidences}. 

The signal and idler photons from each source are separated deterministically using a series of add-drop filters, which acts as a dichroic mirror (`B' in Fig. \ref{fig:GHZ_device}). Paths 1 and 4 carry only signal photons, while paths 2 and 3 carry only idler photons. All four paths carry both logical states, because the ``colour'' and logical state are distinct degrees of freedom in this encoding. At this stage the state can be written as
\begin{align}
     \ket{\psi} =& \left(\ket{\text{vac}} + \left(\beta_1 a^{\dagger(1)}_{S0} a^{\dagger(2)}_{I0} + \beta_2 a^{\dagger(1)}_{S1} a^{\dagger(2)}_{I1} \right)\ket{\text{vac}} \right) \\  \nonumber &\otimes \left(\ket{\text{vac}} + \left(\beta_3 a^{\dagger(4)}_{S0} a^{\dagger(3)}_{I0} + \beta_4 a^{\dagger(4)}_{S1} a^{\dagger(3)}_{I1} \right)\ket{\text{vac}} \right), 
\end{align}
where the superscript on each ladder operator denotes the relevant path (see Fig. \ref{fig:GHZ_device}). Expanding the tensor product results in zero-, two-, and four-photon terms; we restrict our attention to the latter, since four-fold coincidence events will be postselected. The relevant term is 
\begin{align}
    \ket{\psi_{IV}} =& \mathcal{N}\left(\beta_1 a^{\dagger(1)}_{S0} a^{\dagger(2)}_{I0} + \beta_2 a^{\dagger(1)}_{S1} a^{\dagger(2)}_{I1} \right)\ket{\text{vac}}\\ &\otimes \left(\beta_3 a^{\dagger(4)}_{S0} a^{\dagger(3)}_{I0} + \beta_4 a^{\dagger(4)}_{S1} a^{\dagger(3)}_{I1} \right)\ket{\text{vac}} \nonumber,
\end{align}
where $\mathcal{N}$ is a normalization constant. 

Next, paths 1 and 4 are routed to detectors, while paths 2 and 3 are ``merged" at an add-drop filter resonant with $I1$, but not $I0$ (`C' in Fig. \ref{fig:GHZ_device}). The add-drop acts in analogy to a polarizing beamsplitter, effecting the transformation 
\begin{align}
    a^{\dagger(2)}_{I0} &\rightarrow a^{\dagger(2)}_{I0} \nonumber\\
    a^{\dagger(3)}_{I0} &\rightarrow a^{\dagger(3)}_{I0}\nonumber\\
    a^{\dagger(2)}_{I1} &\rightarrow a^{\dagger(3)}_{I1}\nonumber\\
    a^{\dagger(3)}_{I1} &\rightarrow a^{\dagger(2)}_{I1}.
\end{align}
Because we will trace over the ``colour" (S and I), paths 2 and 3 must contain photons of the same colour to obtain a pure GHZ state. The state following the add-drop is
\begin{align}
    \ket{\psi_{IV}} =& \mathcal{N}\left(\beta_1 a^{\dagger(1)}_{S0} a^{\dagger(2)}_{I0} + \beta_2 a^{\dagger(1)}_{S1} a^{\dagger(3)}_{I1} \right)\ket{\text{vac}}\\ &\otimes \left(\beta_3 a^{\dagger(4)}_{S0} a^{\dagger(3)}_{I0} + \beta_4 a^{\dagger(4)}_{S1} a^{\dagger(2)}_{I1} \right)\ket{\text{vac}} \nonumber,
\end{align}
and the state describing four-fold coincidence events is 
\begin{align}
    \ket{\bar{\psi}_{IV}} = \nonumber \bar{\mathcal{N}}\bigg( &\beta_1 \beta_3 a^{\dagger(1)}_{S0} a^{\dagger(2)}_{I0} a^{\dagger(3)}_{I0} a^{\dagger(4)}_{S0}\\ &+ \beta_2 \beta_4 a^{\dagger(1)}_{S1} a^{\dagger(2)}_{I1} a^{\dagger(3)}_{I1} a^{\dagger(4)}_{S1}\bigg)\ket{\text{vac}}\\
    =\bar{\mathcal{N}}\bigg( &\beta_1 \beta_3 \ket{0000}+ \beta_2 \beta_4 \ket{1111}\bigg)\ket{SIIS},
\end{align}
which is a pure frequency-bin-encoded GHZ state after tracing over the colour ($S$ or $I$).


{We envision the source in Fig. \ref{fig:W_device}a to be a pair of silicon microring resonators. Such sources have been shown to generate nearly uncorrelated photon pairs with $|\beta|^2\approx 0.1$ for pump pulse durations around 10 ps, with a 1 MHz repetition rate; this corresponds to a pair rate of $\sim 10^6$ Hz \cite{Grassani2016}. From this we predict the probability of generating a pair of pairs per pump pulse to be $\sim|\beta|^4 \approx 0.01$, giving a four-photon generation rate $\mathcal{R}_{IV} \sim 10^5$ Hz.} Half of these sets of photons lead to four-fold coincidences after the add-drop {(see Appendix \ref{section:AppendixA})}, so we expect a GHZ generation rate of $10^4 - 10^5$ Hz for milliwatt pump powers. 

\section{Three-photon W states}
\label{section:W}

We now turn to the generation of three-photon W states. W states are known to be relatively robust against loss \cite{PhysRevA.62.062314} and, like GHZ states, they have been explored as resources for a number of quantum protocols \cite{Zhu2015, PRA_Agrawal, Yang2017_W_dense_coding}. {A maximally entangled three-qubit W state has the form}
\begin{align}
    \ket{W} = \frac{1}{\sqrt{3}}\left(\ket{100} + \ket{010} + \ket{001} \right). \label{eq:W_def}
\end{align}

\begin{figure}[h]
    \centering
    \includegraphics[width=0.5\textwidth]{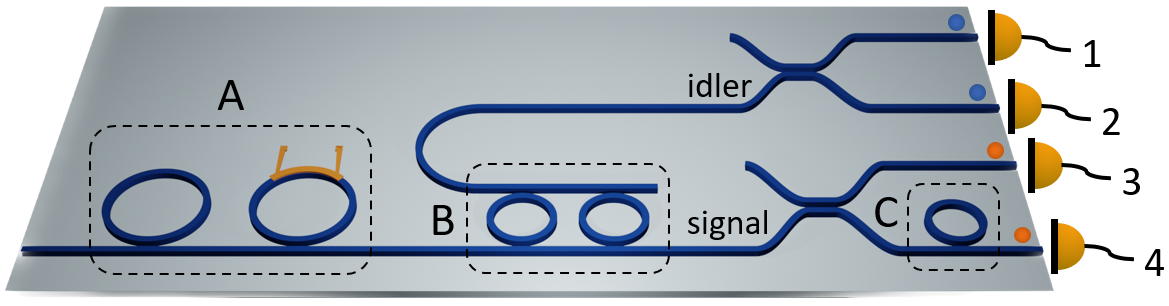}
    \caption{The integrated W state source. Box A labels the photon source. Signal and idler photons are separated at the demultiplexer (B), and routed toward detectors. C labels a filter which transmits only signal photons in frequency bin 1.}
    \label{fig:W_device}
\end{figure}

Our scheme for generating such states begins with the generation of four photons in a single source (labelled `A' in Fig. \ref{fig:W_device}) with the pumping scheme sketched in Fig. \ref{fig:Dual_pumping}. {In this case the two rings' resonances are different, and they can be excited simultaneously through the same bus waveguide and a properly engineered pump.}

 Here 
\begin{align}
    C^{\dagger}_{II} = \frac{1}{\beta}\left\{ \beta_1 a^{\dagger}_{S0} a^{\dagger}_{I1} + \beta_2 a^{\dagger}_{S1} a^{\dagger}_{I0}\right\},
\end{align}
with $\beta_1$ and $\beta_2$ defined in Eqs. \eqref{eq:beta1} and \eqref{eq:beta2}. Inserting this in Eq. \eqref{eq:ket_approx} and restricting our attention to the four-photon terms, we have
\begin{align}
    \ket{\psi_{IV}} &= \mathcal{N}\big( \beta_1^2 a^{\dagger}_{S0} a^{\dagger}_{S0} a^{\dagger}_{I1} a^{\dagger}_{I1} + \beta_2^2 a^{\dagger}_{S1} a^{\dagger}_{S1} a^{\dagger}_{I0} a^{\dagger}_{I0} \nonumber \\ &+ 2\beta_1 \beta_2 a^{\dagger}_{S0} a^{\dagger}_{S1} a^{\dagger}_{I0} a^{\dagger}_{I1}\big)\ket{\text{vac}}. \label{eq:Wsource}
\end{align}

As in the GHZ device, the signal and idler photons are separated deterministically by add-drop filters (`B' in Fig. \ref{fig:W_device}); the signal photons are routed to path 4, and idler photons to path 2. Each path is sent into a directional coupler (DC) which acts as a 50-50 beamsplitter. Three of the four DC outputs lead directly to detectors: These are the three `parties' among which the W state is shared. {The fourth output (path 4) is filtered ('C' in Fig. \ref{fig:W_device}), leaving only photons generated in resonance $S1$, and then routed to a herald detector}. After this processing, a four-photon coincidence event is described by
\begin{align}
    \ket{IV'} =& \mathcal{N}'\bigg[ 4\beta_2^2 \ket{0,0,1,1}\nonumber\\
    & + 2\beta_1 \beta_2 \left(\ket{0,1,0,1} + \ket{1,0,0,1} \right) \bigg]\ket{I,I,S,S},
\end{align}
which is derived in Appendix A. 

Here again we expect the rate of pairs of pairs generated in the source to be $\sim 10^5$ Hz (see Section \ref{section:GHZ}). The fraction of four-photon sets that could lead to four-fold coincidences after the DCs and filtering in path 4 is given by 
\begin{align}
    \mathcal{F} = \frac{1}{8} \left(\frac{2|\beta_2|^4 + |\beta_1|^2 |\beta_2|^2}{|\beta_1|^4 + |\beta_2|^4 + |\beta_1|^2 |\beta_2|^2}\right),
\end{align}
where we have set $\mathcal{R} = \mathcal{T} = \frac{1}{\sqrt{2}}$ for the DCs, which should be configured as 50-50 beamsplitters to maximize the generation rate. The total rate of W states is $\mathcal{R}_{W} = \mathcal{F} \mathcal{R}_{IV}$. The rate depends on the relative amplitudes of $\beta_1$ and $\beta_2$; for example, to generate the W state of Eq. \eqref{eq:W_def}, one would set $\beta_1 = 2\beta_2$, giving $\mathcal{F}=\frac{1}{28}$. Then with a silicon microring source we estimate $\mathcal{R}_{W} \sim 10^3 - 10^4$ Hz for milliwatt pump powers.

We point out that the redundant degree of freedom can be exploited to improve the efficiency: without the signal/idler label, the photons could not be deterministically separated into two multiple paths, so the fraction of photons leading to four-fold coincidences would be smaller. Indeed, in this respect, this scheme is more efficient compared to a similar scheme where the logical states were simply `red' and `blue', with three beamsplitters required to route them to four detectors \cite{Fang_PRL, Menotti_PRA}. 



\section{Conclusion}
\label{section:Conclusions}

Frequency bin encoding has emerged as a convenient approach to the generation of non-classical light, particularly in integrated platforms. The generation of high-fidelity Bell states, for example, has been demonstrated; here we have explored the generation of multipartite states. We have described integrated sources of frequency-bin-encoded W and GHZ states, based on the manipulation and postselection of photon pairs generated by SFWM in pulsed microring sources. For typical silicon microring sources pumped by picojoule pump pulses, we predict generation rates of $10^3 - 10^4$ Hz for both W and GHZ states. The implementation of these schemes is a clear next step in extending frequency bin encoding towards the generation of multipartite entangled states.

Frequency bin encoding can accommodate more than two logical levels, and integrated platforms are well-suited for multiplexing many photon sources; it is natural to consider whether one could conceive sources of more complex, high-dimensional multipartite states. The schemes discussed here cannot be generalized in a straightforward way, a difficulty that has been encountered by others \cite{Erhard2018}. It is challenging to comment more generally on the resources required to generate a particular state, or conversely, about what states can be generated given a particular set of experimental resources. Indeed, it has been demonstrated that one cannot even efficiently compute the high-dimensional state generated by a given experimental scheme \cite{Krenn_PhysRevLett.119.240403}. Yet the extension of this work to the design of more complex sources is an obvious topic for future work, especially given the ease with which the platforms discussed here can be implemented, relative to other scenarios for investigating high-dimensional degrees of freedom.

\begin{acknowledgments}
 M.B. acknowledges support from the University of Toronto Faculty of Arts \& Science Top Doctoral Fellowship. M.L. acknowledges support by PNRR MUR project PE0000023-NQSTI. J.E.S. and M.B. acknowledge support from the Natural Sciences and Engineering Research Council of Canada. 
\end{acknowledgments}

\bibliography{apssamp}

\onecolumngrid
\appendix
\pagebreak

\section{W and GHZ details}

\label{section:AppendixA}

Here we provide a more detailed description of the W and GHZ generation schemes. 

\subsection{Three-photon W state}
We consider the W state device (Fig. \ref{fig:W_device}) with the photon pair source driven in the dual-pump scheme (Fig. \ref{fig:Dual_pumping}). Assuming that the generated photon pairs are roughly separable, the nonlinear Hamiltonian describing the production of pairs can be taken to be
\begin{align}
    H_{NL} = \hbar \Gamma \alpha^2(t) a^{\dagger}_{S0} a^{\dagger}_{I1} + \hbar \Gamma' \alpha'^2(t) a^{\dagger}_{S1} a^{\dagger}_{I0} + H.c.
\end{align}
Here $\Gamma$ and $\Gamma'$ are nonlinear coupling rates associated with each ring, $\alpha$ and $\beta$ are classical pump amplitudes in the two rings, and $a^{\dagger}_{J}$ are raising operators associated with a ring resonance $J$. Neglecting time ordering corrections, this generates the state
\begin{align}
    \ket{\psi} &= e^{-\frac{i}{\hbar}\int^t_{t_0} dt' H_{NL}(t')} \ket{\text{vac}}\\
    &= e^{-{i}\int^t_{t_0} dt' \left(\Gamma \alpha(t') ^2 a^{\dagger}_{S0} a^{\dagger}_{I1} + \Gamma' \alpha'^2(t') a^{\dagger}_{S1} a^{\dagger}_{I0} + H.c.\right)} \ket{\text{vac}} \\
    &= e^{\left\{-{i}\Gamma \left(\int^t_{t_0} dt'\alpha^2(t') \right) a^{\dagger}_{S0} a^{\dagger}_{I1} -{i}\Gamma' \left(\int^t_{t_0} dt'\alpha'^2(t') \right) a^{\dagger}_{S1} a^{\dagger}_{I0}\right\} - H.c. } \ket{\text{vac}}\\
    &\equiv e^{\beta C^{\dagger}_{II} - H.c.}\ket{\text{vac}},
    \label{eq:state_1}
\end{align}
where we introduce the pair generation operator
\begin{align}
    C^{\dagger}_{II} &= \frac{1}{\beta}\left\{ -{i}\Gamma \left(\int^t_{t_0} dt'\alpha^2(t') \right) a^{\dagger}_{S0} a^{\dagger}_{I1} -{i}\Gamma' \left(\int^t_{t_0} dt'\alpha'^2(t') \right) a^{\dagger}_{S1} a^{\dagger}_{I0}\right\} \\
    &\equiv  \frac{1}{\beta}\left\{ \beta_1 a^{\dagger}_{S0} a^{\dagger}_{I1} + \beta_2 a^{\dagger}_{S1} a^{\dagger}_{I0}\right\}. 
\end{align}
Here $|\beta|^2 = {|\beta_1|^2 + |\beta_2|^2}$ is the probability of generating a pair per pump pulse, and $C^{\dagger}_{II}\ket{\text{vac}}$ is a normalized two photon state.

We consider the regime of low pair generation probability, such that pairs of photons and pairs of pairs are generated, with negligible higher-order events. The state generated by the source then is approximately 
\begin{align}
    \ket{\psi} &= \left(1+\mathcal{O}(\beta^2)\right)\ket{\text{vac}} + \beta C^{\dagger}_{II}\ket{\text{vac}}+\frac{\beta^2}{2}C^{\dagger2}_{II} \ket{\text{vac}}.
    \label{eq:source1}
\end{align}
which is just Eq. \eqref{eq:state_1} to second order. If we set $\beta_2 = \beta_1 e^{i\phi}$, then the two-photon term above (when normalized) corresponds to a frequency-bin-encoded Bell state of the form 
\begin{align}
    \ket{\Psi} &= \frac{1}{\sqrt{2}} \left( a^{\dagger}_{S0} a^{\dagger}_{I1} + e^{i\phi} a^{\dagger}_{S1} a^{\dagger}_{I0} \right) \ket{\text{vac}}\\
    &= \frac{1}{\sqrt{2}}\left( \ket{0 1} + e^{i\phi} \ket{1 0}\right), 
\end{align}
where the two qubits are the signal and idler photons, and their logical states are encoded in the frequency bins in which they are generated.

Here we are interested in the four-photon terms that arise from the latter term in \eqref{eq:source1}. Inserting $C^{\dagger}_{II}$ we have 
\begin{align}
    \ket{\psi_4} = \mathcal{N}\left( \beta_1^2 a^{\dagger}_{S0} a^{\dagger}_{S0} a^{\dagger}_{I1} a^{\dagger}_{I1} + \beta_2^2 a^{\dagger}_{S1} a^{\dagger}_{S1} a^{\dagger}_{I0} a^{\dagger}_{I0} + 2\beta_1 \beta_2 a^{\dagger}_{S0} a^{\dagger}_{S1} a^{\dagger}_{I0} a^{\dagger}_{I1}\right)\ket{\text{vac}}, \label{eq:source}
\end{align}
where 
\begin{align}
    \mathcal{N} = \frac{1}{\sqrt{4|\beta_1|^4 + 4|\beta_2|^4 + 4|\beta_1|^2 |\beta_2|^2}} \label{eq:N}
\end{align}
is a normalization constant.

\subsubsection{Routing photons}

We consider a setup where the signal and idler photons generated by the source are split between two paths. Each path is then split by a beamsplitter, and we postselect on four-photon coincidences (see Fig. \ref{fig:W_device}). We can write the operators in Eq. \eqref{eq:source} in terms of operators referring to each detection arm using beamsplitter input-output relations. Recall for a single beamsplitter, where we label the input ports 1 and 2, and the output ports 3 and 4, we have 
\begin{align}
    \begin{bmatrix}
    a_3\\ a_4
    \end{bmatrix} = 
    \begin{bmatrix}
    \mathcal{T} & \mathcal{R} \\
    \mathcal{R} & \mathcal{T}
    \end{bmatrix}
    \begin{bmatrix}
    a_1\\ a_2
    \end{bmatrix},
\end{align}
Where the matrix is unitary. Using this we obtain
\begin{align}
    a_1 &= \mathcal{T}^* a_3 + \mathcal{R}^* a_4 \nonumber\\
    a_2 &= \mathcal{R}^* a_3 + \mathcal{T}^* a_4.
\end{align}
So for our setup we can write 
\begin{align}
    a_I &= \mathcal{R}^*_1 c_I^{(1)} + \mathcal{T}^*_1 c_I^{(2)} \nonumber\\
    a_S &= \mathcal{R}^*_2 c_S^{(3)} + \mathcal{T}^*_2 c_S^{(4)} \label{eq:input-output}
\end{align}
Here $c_J^{(n)}$ are ladder operators associated with the output branch $n$ (see Fig. \ref{fig:W_device}). We now substitute \eqref{eq:input-output} into \eqref{eq:source}. 

The resulting expression involves operators of the form
\begin{align}
    \mathcal{O}_{J,J',K,K'} = a^{\dagger}_{J} a^{\dagger}_{J'} a^{\dagger}_{K} a^{\dagger}_{K'} \label{eq:O_op}
\end{align}
acting on the vacuum, where $J,J'$ are associated with signal photons and $K,K'$ with idler photons. Inserting \eqref{eq:input-output} into \eqref{eq:O_op} we have 
\begin{align}
    \mathcal{O}_{J,J',K,K'} = & \left(\mathcal{R}_2c_{J}^{(3)\dagger}+\mathcal{T}_2c_{J}^{(4)\dagger}\right) \left(\mathcal{R}_2c_{J'}^{(3)\dagger}+\mathcal{T}_2c_{J'}^{(4)\dagger}\right) \nonumber\\
    &\times \left(\mathcal{R}_1c_{K}^{(1)\dagger}+\mathcal{T}_1c_{K}^{(2)\dagger}\right) \left(\mathcal{R}_1c_{K'}^{(1)\dagger}+\mathcal{T}_1c_{K'}^{(2)\dagger}\right) \nonumber
\end{align}
Expanding this we have 
\begin{align}
    {\mathcal{O}}_{J,J',K,K'} &= \mathcal{R}_1 \mathcal{T}_1 \mathcal{R}_2 \mathcal{T}_2 \left( c_{J'}^{(3)\dagger} c_{J}^{(4)\dagger} + c_{J}^{(3)\dagger}c_{J'}^{(4)\dagger}\right)
    \left( c_{K}^{(1)\dagger} c_{K'}^{(2)\dagger} + c_{K'}^{(1)\dagger}c_{K}^{(2)\dagger}\right) + \overline{\mathcal{O}}_{J,J',K,K'} \\ \nonumber
    &=\mathcal{R}_1 \mathcal{R}_2 \mathcal{T}_1 \mathcal{T}_2 \left(c_{J'}^{(3)\dagger} c_{J}^{(4)\dagger}c_{K}^{(1)\dagger} c_{K'}^{(2)\dagger} + c_{J'}^{(3)\dagger}c_{J}^{(4)\dagger}c_{K'}^{(1)\dagger} c_{K}^{(2)\dagger} + c_{J}^{(3)\dagger} c_{J'}^{(4)\dagger}c_{K}^{(1)\dagger} c_{K'}^{(2)\dagger} + c_{J}^{(3)\dagger}c_{J'}^{(4)\dagger}c_{K'}^{(1)\dagger} c_{K}^{(2)\dagger}\right) \\ \nonumber &+ \overline{\mathcal{O}}_{J,J',K,K'}\\
    &=\mathcal{R}_1 \mathcal{R}_2 \mathcal{T}_1 \mathcal{T}_2 \left(c_{K}^{(1)\dagger} c_{K'}^{(2)\dagger} c_{J'}^{(3)\dagger} c_{J}^{(4)\dagger} + c_{K'}^{(1)\dagger} c_{K}^{(2)\dagger}c_{J'}^{(3)\dagger}c_{J}^{(4)\dagger} + c_{K}^{(1)\dagger} c_{K'}^{(2)\dagger}c_{J}^{(3)\dagger} c_{J'}^{(4)\dagger} + c_{K'}^{(1)\dagger} c_{K}^{(2)\dagger}c_{J}^{(3)\dagger}c_{J'}^{(4)\dagger}\right) \nonumber \\  &+ \overline{\mathcal{O}}_{J,J',K,K'},
\end{align}
where we have grouped the terms that cannot result in a four-fold coincidence in the term $\overline{\mathcal{O}}_{J,J',K,K'}$.
Note that if $J=J'$ and $K=K'$, we have
\begin{align}
    {\mathcal{O}}_{J,J,K,K}
    &=4\mathcal{R}_1 \mathcal{R}_2 \mathcal{T}_1 \mathcal{T}_2 \left(c_{K}^{(1)\dagger}c_{K}^{(2)\dagger}c_{J}^{(3)\dagger} c_{J}^{(4)\dagger}\right) + \overline{\mathcal{O}}_{J,J,K,K}
\end{align}

Our notation for the states generated by these operators will be the following:
\begin{align}
    c^{(1)\dagger}_{K} c^{(2)\dagger}_{K'} c^{(3)\dagger}_{J} c^{(4)\dagger}_{J'}\ket{\text{vac}} = \ket{K,K',J,J'}.
\end{align}
We can then see that
\begin{align}
    {\mathcal{O}}_{J,J',K,K'} \ket{\text{vac}} &= \mathcal{R}_1 \mathcal{R}_2 \mathcal{T}_1 \mathcal{T}_2\left( \ket{K,K',J',J} + \ket{K',K,J',J} + \ket{K,K',J,J'} + \ket{K',K,J,J'}\right) + \overline{\mathcal{O}}_{J,J',K,K'} \ket{\text{vac}},
    \label{eq:Ovac}
\end{align}
where the latter term will be excluded by postselecting on four-fold coincidences.

\subsubsection{Output state and filtering}
Recall Eq. \eqref{eq:source}, which is the four-photon state generated by the source. After propagating through the beamsplitters, the part of the state with a photon in each of the four outputs can be written as
\begin{align}
    \ket{IV} = \mathcal{N}\left(\beta_1^2 {\mathcal{O}}_{S0,S0,I1,I1} + \beta_2^2 {\mathcal{O}}_{S1,S1,I0,I0} + 2 \beta_1 \beta_2 {\mathcal{O}}_{S0,S1,I0,I1}\right)\ket{\text{vac}}, 
\end{align}
with $\mathcal{N}$ defined in \eqref{eq:N}. Using \eqref{eq:Ovac} we have
\begin{align}
    \ket{IV} =& \mathcal{N}\bigg(\mathcal{R}_1 \mathcal{R}_2 \mathcal{T}_1 \mathcal{T}_2 \bigg[ 4\beta_1^2\ket{I1,I1,S0,S0} + 4\beta_2^2 \ket{I0,I0,S1,S1}\nonumber+ 2\beta_1 \beta_2\big(\ket{I0,I1,S1,S0} + \ket{I1,I0,S1,S0}\\
    &+ \ket{I0,I1,S0,S1} + \ket{I1,I0,S0,S1} \big) \bigg] + \bigg[ \beta_1^2 {\mathcal{O}}_{S0,S0,I1,I1} + \beta_2^2 {\mathcal{O}}_{S1,S1,I0,I0} + 2 \beta_1 \beta_2 {\mathcal{O}}_{S0,S1,I0,I1} \bigg]\ket{\text{vac}}\bigg), \label{eq:four-photon_full}
\end{align}
and we define
\begin{align}
    \ket{IV'} =& \mathcal{N}'\bigg[ 4\beta_1^2\ket{I1,I1,S0,S0} + 4\beta_2^2 \ket{I0,I0,S1,S1}\nonumber\\
    & + 2\beta_1 \beta_2\left(\ket{I0,I1,S1,S0} + \ket{I1,I0,S1,S0} + \ket{I0,I1,S0,S1} + \ket{I1,I0,S0,S1} \right) \bigg],
\end{align}
where $\ket{IV'}$ is the state describing four-fold coincidence events only. We separate the `colour' (signal/idler) and frequency bin  (0/1) degrees of freedom by writing, e.g., $\ket{S0}$ explicitly as a composite system with two degrees of freedom $\ket{S}\ket{0}$. In this notation we have 
\begin{align}
    \ket{IV'} =& \mathcal{N}'\bigg[ 4\beta_1^2 \ket{1,1,0,0} + 4\beta_2^2 \ket{0,0,1,1}\nonumber\\
    & + 2\beta_1 \beta_2 \left(\ket{0,1,1,0} + \ket{1,0,1,0} + \ket{0,1,0,1} + \ket{1,0,0,1} \right) \bigg]\ket{I,I,S,S},
\end{align}
and clearly if we trace over the `colour' we have a pure state in the frequency bin degree of freedom. Finally, we also postselect on frequency bin 1 in the herald detector, which we take to be detector 4. Doing this we are left with 
\begin{align}
    \ket{IV'} =& \mathcal{N}''\bigg[ 4\beta_2^2 \ket{0,0,1}_{123}
     + 2\beta_1 \beta_2 \left(\ket{0,1,0}_{123}  + \ket{1,0,0}_{123} \right) \bigg]\ket{1}_4
     \label{eq:W_general},\\
     \mathcal{N}'' =& \frac{1}{\sqrt{16|\beta_2|^4 + 8|\beta_1|^2|\beta_2|^2}}.
\end{align}
Now if we set the pump powers such that $\beta_1 \beta_2 = 2 \beta_1^2$, we have 
\begin{align}
    \ket{IV'} = \frac{1}{\sqrt{3}} \left( \ket{0,0,1} +
    \ket{0,1,0} + \ket{1,0,0}
    \right)_{123} \ket{1}_4, \label{eq:prototypical_W}
\end{align}
which is a W state in ports 1,2,3. Notice one could also use the pump phases to add a relative phase to one of the three terms, but not arbitrary relative phases between the three terms; in principle it seems this could be done by adding a frequency-dependent phase shift to one of the paths. Of course, the relative amplitudes between the terms can be modified; although the state is not completely general, the tunability in \eqref{eq:W_general} is sufficient to construct a W state that is suitable 
for superdense coding and perfect teleportation \cite{PhysRevA.74.062320}.

\subsubsection{GHZ rate}

Many of the photon pair pairs generated by the microring source do not result in four-fold coincidences. We can find the fraction of photons that remain by computing the overlap between $\ket{IV'}$ in \eqref{eq:W_general}, which is the post-selected state, and $\ket{IV}$ in \eqref{eq:four-photon_full}, which is the full four-photon state. We find 
\begin{align}
    |\braket{IV|IV'}|^2 &= |\mathcal{N}|^2|\mathcal{N}''|^2 |\mathcal{R}_1 \mathcal{R}_2\mathcal{T}_1
    \mathcal{T}_2|^2 \left(16 |\beta_2|^4 + 8|\beta_1|^2 |\beta_2|^2 \right)^2\\
    &= \left(\frac{1}{{4|\beta_1|^4 + 4|\beta_2|^4 + 4|\beta_1|^2 |\beta_2|^2}}\right) \left(\frac{1}{{16|\beta_2|^4 + 8|\beta_1|^2|\beta_2|^2}}\right)|\mathcal{R}_1 \mathcal{R}_2\mathcal{T}_1
    \mathcal{T}_2|^2 \left(16 |\beta_2|^4 + 8|\beta_1|^2 |\beta_2|^2 \right)^2.
\end{align}
Putting $\mathcal{R}_1=\mathcal{R}_2 = \mathcal{T}_1 = \mathcal{T}_2 = \frac{1}{\sqrt{2}}$, which optimizes rate of W states, we have  
\begin{align}
    |\braket{IV|IV'}|^2 
    &= \left(\frac{16 |\beta_2|^4 + 8|\beta_1|^2 |\beta_2|^2 }{{4|\beta_1|^4 + 4|\beta_2|^4 + 4|\beta_1|^2 |\beta_2|^2}}\right) \frac{1}{16}\\
    &= \left(\frac{2|\beta_2|^4 + |\beta_1|^2 |\beta_2|^2 }{{|\beta_1|^4 + |\beta_2|^4 + |\beta_1|^2 |\beta_2|^2}}\right) \frac{1}{8}.
\end{align}
If we put $2\beta_2 = \beta_1$, which gives the W state of Eq. \eqref{eq:prototypical_W}, we have
\begin{align}
    |\braket{IV|IV'}|^2 
    &= \left(\frac{6}{{21}}\right) \frac{1}{8} = \frac{1}{28}.
\end{align}
That is, of the pairs of pairs generated in the source, $\frac{1}{28}$ lead to four-fold coincidences; we expect
\begin{align}
    \mathcal{R}_{W} = \mathcal{R}_{IV}/28,
\end{align}
where $\mathcal{R}_{W}$ is the rate of W states and $\mathcal{R}_{IV}$ is the rate of photon-pair-pairs from the source.

\subsection{Four-photon GHZ state}

\subsubsection{The sources}

Here we take two SFWM sources in the pair generation regime, with the resonances configured as shown in Fig. \ref{fig:single_pump}. The nonlinear Hamiltonian describing each source can be taken to be
\begin{align}
    H_{NL}(t) = \hbar \Gamma \alpha(t) a^{\dagger}_{S0} a^{\dagger}_{I0} + \hbar \Gamma' \alpha'(t) a^{\dagger}_{S1} a^{\dagger}_{I1} + H.c.
\end{align}
Neglecting time ordering corrections, the state generated by this Hamiltonian is
\begin{align}
    \ket{\psi} &= e^{-\frac{i}{\hbar}\int^t_{t_0} dt' H_{NL}(t')} \ket{\text{vac}}\\
    &= e^{-{i}\int^t_{t_0} dt' \left(\Gamma \alpha(t') a^{\dagger}_{S0} a^{\dagger}_{I0} + \Gamma' \alpha'(t') a^{\dagger}_{S1} a^{\dagger}_{I1} + H.c.\right)} \ket{\text{vac}} \\
    &= e^{\left\{-{i}\Gamma \left(\int^t_{t_0} dt'\alpha(t') \right) a^{\dagger}_{S0} a^{\dagger}_{I0} -{i}\Gamma \left(\int^t_{t_0} dt'\alpha'(t') \right) a^{\dagger}_{S1} a^{\dagger}_{I1}\right\} - H.c. } \ket{\text{vac}}\\
    &\equiv e^{\beta C^{\dagger}_{II} - H.c.}\ket{\text{vac}},
    \label{eq:state}
\end{align}
where now
\begin{align}
    C^{\dagger}_{II} &= \frac{1}{\beta}\left\{ -{i}\Gamma \left(\int^t_{t_0} dt'\alpha(t') \right) a^{\dagger}_{S0} a^{\dagger}_{I0} -{i}\Gamma \left(\int^t_{t_0} dt'\alpha'(t') \right) a^{\dagger}_{S1} a^{\dagger}_{I1}\right\} \\
    &\equiv  \frac{1}{\beta}\left\{ \beta_1 a^{\dagger}_{S0} a^{\dagger}_{I0} + \beta_2 a^{\dagger}_{S1} a^{\dagger}_{I1}\right\}. 
\end{align}
Here $|\beta|^2 = {|\beta_1|^2 + |\beta_2|^2}$ is the probability of generating a pair per pump pulse, and $C^{\dagger}_{II}\ket{\text{vac}}$ is a normalized two photon state. If we set $\beta_2 = \beta_1 e^{i\phi}$ then the two photon state corresponds to a frequency-bin-encoded Bell state of the form 
\begin{align}
    \ket{\Phi} &= \frac{1}{\sqrt{2}} \left( a^{\dagger}_{S0} a^{\dagger}_{I0} + e^{i\phi} a^{\dagger}_{S1} a^{\dagger}_{I1} \right) \ket{\text{vac}}\\
    &= \frac{1}{\sqrt{2}}\left( \ket{0 0} + e^{i\phi} \ket{1 1}\right),
\end{align}
where the two qubits are the signal and idler photons, and their logical states are encoded in the frequency bins in which they are generated.

Returning the more general case, we consider the low pair generation probability regime, such that the state generated by each source is approximately 
\begin{align}
    \ket{\psi} &= \ket{\text{vac}} + \beta C^{\dagger}_{II}\ket{\text{vac}}\\
    &= \ket{\text{vac}} + \left(\beta_1 a^{\dagger}_{S0} a^{\dagger}_{I0} + \beta_2 a^{\dagger}_{S1} a^{\dagger}_{I1} \right)\ket{\text{vac}} \label{eq:source2}.
\end{align}
We neglect four-photon term in Eq. \eqref{eq:source2} because such terms will not result in four-fold coincidences. 

\subsubsection{Manipulation}

We now discuss the manipulation of the photons in the GHZ device (see Fig. \ref{fig:GHZ_device}). We begin with two photon pair sources configured as described above (see Eq. \eqref{eq:source2}). We label the two spatial modes associated with the sources path 1 and 4. The state generated by the two rings is approximately 
\begin{align}
    \ket{\psi} &= \left(\ket{\text{vac}} + \left(\beta_1 a^{\dagger(1)}_{S0} a^{\dagger(1)}_{I0} + \beta_2 a^{\dagger(1)}_{S1} a^{\dagger(1)}_{I1} \right)\ket{\text{vac}} \right) \otimes \left(\ket{\text{vac}} + \left(\beta_3 a^{\dagger(4)}_{S0} a^{\dagger(4)}_{I0} + \beta_4 a^{\dagger(4)}_{S1} a^{\dagger(4)}_{I1} \right)\ket{\text{vac}} \right) \\
    &= \ket{\psi_{0,2}} + \left(\beta_1 a^{\dagger(1)}_{S0} a^{\dagger(1)}_{I0} + \beta_2 a^{\dagger(1)}_{S1} a^{\dagger(1)}_{I1} \right)\ket{\text{vac}} \otimes \left(\beta_3 a^{\dagger(4)}_{S0} a^{\dagger(4)}_{I0} + \beta_4 a^{\dagger(4)}_{S1} a^{\dagger(4)}_{I1} \right)\ket{\text{vac}}, \label{eq:tensor_prod}
\end{align}
where in \eqref{eq:tensor_prod} we have grouped the terms that result in the generation of no photons or photon pairs in $\ket{\psi_{0,2}}$. After each source, the signal and idler photons are separated using an add-drop filter; the path taken by signal photons remains unchanged, while idler photons are routed into a new path. The state is then 
\begin{align}
    \ket{\psi}&=\ket{\psi_{0,2}} + \left(\beta_1 a^{\dagger(1)}_{S0} a^{\dagger(2)}_{I0} + \beta_2 a^{\dagger(1)}_{S1} a^{\dagger(2)}_{I1} \right)\ket{\text{vac}} \otimes \left(\beta_3 a^{\dagger(4)}_{S0} a^{\dagger(3)}_{I0} + \beta_4 a^{\dagger(4)}_{S1} a^{\dagger(3)}_{I1} \right)\ket{\text{vac}}.
\end{align}
Next, paths 2 and 3 are mixed at an add-drop filter. The ring is resonant with the frequency bin $I1$, so photons in bin $I0$ remain in the same path while photons in $I1$ are swapped. That is, the add-drop effects the transformation 
\begin{align}
    a^{\dagger(2)}_{I0} \rightarrow a^{\dagger(2)}_{I0}\\
    a^{\dagger(3)}_{I0} \rightarrow a^{\dagger(3)}_{I0}\\
    a^{\dagger(2)}_{I1} \rightarrow a^{\dagger(3)}_{I1}\\
    a^{\dagger(3)}_{I1} \rightarrow a^{\dagger(2)}_{I1},
\end{align}
so the state becomes 
\begin{align}
    \ket{\psi}&=\ket{\psi_{0,2}} + \left(\beta_1 a^{\dagger(1)}_{S0} a^{\dagger(2)}_{I0} + \beta_2 a^{\dagger(1)}_{S1} a^{\dagger(3)}_{I1} \right)\ket{\text{vac}} \otimes \left(\beta_3 a^{\dagger(4)}_{S0} a^{\dagger(3)}_{I0} + \beta_4 a^{\dagger(4)}_{S1} a^{\dagger(2)}_{I1} \right)\ket{\text{vac}}.
\end{align}
Expanding the tensor product we have
\begin{align}
    \ket{\psi}=\ket{\psi_{0,2}} &+ \left(
    \beta_1 \beta_4
    a^{\dagger(1)}_{S0} a^{\dagger(2)}_{I0} a^{\dagger(4)}_{S1} a^{\dagger(2)}_{I1}
    + \beta_2 \beta_3
    a^{\dagger(1)}_{S1} a^{\dagger(3)}_{I1}
    a^{\dagger(4)}_{S0} a^{\dagger(3)}_{I0}
    \right)\ket{\text{vac}}
    \nonumber \\
    &+ \left(\beta_1 \beta _3  a^{\dagger(1)}_{S0} a^{\dagger(2)}_{I0} a^{\dagger(4)}_{S0} a^{\dagger(3)}_{I0}
    + \beta_2 \beta_4
    a^{\dagger(1)}_{S1} a^{\dagger(3)}_{I1}
    a^{\dagger(4)}_{S1} a^{\dagger(2)}_{I1}
    \right)\ket{\text{vac}}.
\end{align}
Only the third term can lead to four-fold coincidences, while the first two cannot. We focus on the last term
\begin{align}
    \ket{GHZ} &= \mathcal{N} \left(\beta_1 \beta _3 a^{\dagger(1)}_{S0} a^{\dagger(2)}_{I0} 
    a^{\dagger(3)}_{I0}
    a^{\dagger(4)}_{S0} 
    + \beta_2 \beta_4
    a^{\dagger(1)}_{S1}
    a^{\dagger(2)}_{I1}
    a^{\dagger(3)}_{I1}
    a^{\dagger(4)}_{S1} 
    \right)\ket{\text{vac}}\\
    &= \mathcal{N} \bigg(\beta_1 \beta _3 \ket{0000}\ket{SIIS} 
    + \beta_2 \beta_4 \ket{1111}\ket{SIIS} 
    \bigg)\\
    &= \mathcal{N} \bigg(\beta_1 \beta _3 \ket{0000}
    + \beta_2 \beta_4 \ket{1111}
    \bigg),
\end{align}
where we have introduced the same notation used above to distinguish the frequency bin and `colour' degrees of freedom, and traced over the latter. We have a four-photon GHZ state with arbitrary relative amplitude and phase between the two terms.

\end{document}